\documentclass[preprint,prd,aps,superscriptaddress,nofootinbib]{revtex4}
\usepackage{graphicx}

\begin{document}

\title {\large Possible Effects of Dark Energy on the Detection
of Dark Matter Particles}

\author{Peihong Gu}
\affiliation{ Institute of High Energy Physics, Chinese Academy of
Sciences, P.O. Box 918-4, Beijing 100049, P. R. China}

\author{Xiao-Jun Bi }
\affiliation{ Key laboratory of particle astrophysics,
Institute of High Energy Physics, Chinese Academy of
Sciences, P.O. Box 918-3, Beijing 100049, P. R. China}

\author{Zhi-Hai Lin}
\affiliation{ Institute of High Energy Physics, Chinese Academy of
Sciences, P.O. Box 918-4, Beijing 100049, P. R. China}

\author{Xinmin Zhang}
\affiliation{ Institute of High Energy Physics, Chinese Academy of
Sciences, P.O. Box 918-4, Beijing 100049, P. R. China}

\date{\today}

\begin{abstract}

We study in this paper the possible influence of the dark energy on the
detection of the
dark matter particles. In models of dark energy described by a
dynamical scalar field such as the Quintessence, its interaction with the
dark matter will cause the dark matter particles such as the
neutralino vary as a function of space and time. Given a specific model of
the Quintessence and its interaction in this paper we
calculate  numerically the corrections to the neutralino masses
and
the induced spectrum of the neutrinos from the annihilation of
the neutralinos pairs in the
core of the Sun.
This study gives rise to a possibility of probing for dark
energy in the experiments of detecting the dark matter particles.

\end{abstract}

\maketitle


Recent observational data from supernovae (SN) Ia \cite{sn} and cosmic
microwave
background radiation (CMBR) \cite{cmb} strongly support for
the `cosmic concordance' model, in which the
Universe is spatially flat with ~4\% baryon matter, ~23\% of
cold dark matter (DM) and ~73\% of dark energy (DE).
The baryon matter is well described by the standard model of the particle
physics, however the nature of the dark matter and the dark energy remains
unknown.

There have been many proposals in the literature for the dark
matter candidates theoretically.  From the point of
view of the particle physics, the leading candidates for cold dark matter
are
the axion and the neutralino. Various experiments in the search
directly or indirectly for these
dark matter particles are currently under way.

Regarding dark energy, the simplest candidate
seems to be a remnant small cosmological constant. However,
many physicists are attracted by
the idea that dark energy is due to a dynamical component, such as a
canonical scalar field $Q$,
named {\it Quintessence} \cite{quin}.
Being a dynamical component, the scalar field of the dark energy is
expected to interact with the other matters \cite{int}.
There are many discussions on the explicit couplings of
quintessence to baryons, dark matter, photons and neutrinos. These
interactions if exist will open up the possibilities of probing
non-gravitationally for the dark
energy. In this paper we consider the possible effects of the
dark energy models which interact with the dark matter
in the detection of the dark matter particles. Specifically we will study the
influence of the dark energy on the neutralino masses in the
Sun, and then calculate the neutrino spectrum annihilated
from the neutralino pairs in the core of the Sun.

We start with a coupled system of
the interacting dark energy and the dark matter with
the Lagrangian generally given by
\begin{eqnarray}
\mathcal{L}_{eff}=\mathcal{L}_{\rm DM}
+\mathcal{L}_{\phi}+\mathcal{L}_{int},
\end{eqnarray}
where $\mathcal{L}_{\rm DM}$ and $\mathcal{L}_{\phi}$ are the
free Lagrangian for dark matter and dark energy and the interaction part
is given by
\begin{eqnarray}
\mathcal{L}_{int}&=&
-\frac{1}{2}M_{\chi}(\phi){\bar \chi}{\chi}
-M_{\psi}(\phi){\bar \psi}{\psi}
-\frac{1}{2}M_S^2(\phi) S^2
-\frac{g_{\chi}}{\Lambda}
 \partial_\mu \phi {\bar \chi} \gamma^\mu \gamma^5 \chi
\nonumber \\
& &
-\frac{g_{\psi}}{\Lambda}
 \partial_\mu \phi {\bar \psi} \gamma^\mu  \psi
- \frac{g_S}{\Lambda}\partial_\mu \phi S  \partial^\mu S
-\frac{g_S^\prime}{\Lambda}  \phi\partial_\mu \phi  \partial^\mu S
- \sum \frac{g_i}{\Lambda^2} \mathcal{O}_i \ \ ,
\end{eqnarray}
where the dark matter particles could be the boson($S$), Majorana
fermion($\chi$) and(or) Dirac fermion($\psi$). In the detailed
discussions below we will consider only the Majorana fermion such
as the neutralino to be the dark matter particle and focus on the
interactions which affect the dark matter particles via the mass
terms.

Being a function of the
quintessence scalar $\phi$ the mass of the dark matter particle will vary
during the evolution of the universe. As shown in Refs.
\cite{riotto,rosenfeld,bi2} this helps solve the
coincidence problem. Furthermore,
 this type of interactions
will affect the cosmic structure formation \cite{peebles},
and the power spectrum of CMB \cite{hof}.
In this paper we will present a new possible effect of the interacting dark
energy with the dark matter in the detection
of the dark matter particles.

There are in general two different ways, direct and indirect, in the
detections of the dark matter particles.
 The direct detection records the recoil energy of the
detector nuclei when the dark matter particles scatter off them as they
pass through
the Earth and interact with the matter. The indirect detection observes
the annihilation products by the dark matter particles.
Obviously the expected detection rates depend on the mass of the dark
matter particles. In the presence of the interaction the mass of the dark
matter particle will vary as a function of time and also space,
and consequently the dark energy will influence the detection.

To determine the mass of neutralino as a function of space we need to
know the value of the dark energy scalar field as a function of space.
Taking into account
the back reaction of the interaction between the dark matter and the dark
energy the effective potential of the dark energy scalar as a function of
the energy density of the cold dark matter $\rho_{\chi}(\phi)$ is given by
\begin{equation}
\label{veff1}
V_{eff}=\rho_{\chi}(\phi)+V(\phi).
\end{equation}
For different dark matter densities the values of dark energy scalar
field are expected to be
different, and consequently the mass of dark matter particles will also be
different. For example,
the mass of the dark matter particle in the center of the Milky Way could be
different from
that in the halo near the solar system. Therefore, the
gamma ray spectrum, or the
synchrotron radiation spectrum, from the galactic center could be different
from that in the nearby halo.
Especially for the neutralino dark matter particle
its mass measured
at the future linear collider (LC) or the large hadron collider (LHC)
on the Earth may be different from that measured in
other places in the Milky Way, such as in the galactic center.
Similarly, the spectrum of the dark matter radiation in
the Milky Way might also be different from other galaxies in the
local group.

In the following we will consider a specific model of the Quintessence and its
interaction with the dark matter particle, the neutralino, and then study its
effects on the indirect detection via the process of the neutralinos
annihilation into neutrinos.

The dark energy potential which we take is
\begin{equation}
\label{poten}
V(\phi)=V_{0}e^{\beta \phi /m_{p}},
\end{equation}
and the interaction between the dark energy and the dark matter
particle are given by
\begin{equation}
\label{dmp}
M_{\chi}(\phi)=M_{\chi_{0}}\left(1+\frac{\lambda_\chi\phi}{m_p}\right)\
.
\end{equation}
Here $m_p$ is the reduced Plank mass, $m_{p}=2.436 \times 10^{18}$
GeV. We have numerically solved the evolution of the cosmological
model with Quintessence potential (\ref{veff1}). And our results
show that the cosmological observations are satisfied with a
suitable choice of the model parameters such as $V_0=4.2\times
10^{-47}{\rm GeV}^4$, $\beta= 1 $ and $\lambda_\chi= 0.1$.

In additional to the interactions between the dark matter and the dark
energy one expects also a coupling of the Quintessence scalar to the
baryon
\footnote{The interactions between the dark energy scalar and the
neutrinos have been considered in Refs. \cite{hong1,gu,nelson,hong2,kaplan,
peccei,zhang} and related implications of mass varying neutrinos
have been studied in neutrino oscillations \cite{kaplan,guen,barger}
and gamma ray burst \cite{zhang}.} via
for instance the quantum gravity effects
\begin{eqnarray}\label{phiqq}
\mathcal{L}_{\phi qq}=-\lambda_B  \frac{\phi }{m_{\rm p}}
 (y_q\bar{Q}_LH q_R)\ .
\end{eqnarray}
The parameter $\lambda_B$ above characterizing the strength of this type of
interaction will be shown below to be strongly constrained. However,
since the baryon density inside the Sun is much higher than any other
matter
densities, this interaction in Eq. (\ref{phiqq}) will be important to our
study in this paper as well as that on the neutrino oscillations
\cite{kaplan,guen,barger}.

The Eq. (\ref{dmp}) shows that the mass of the dark matter particles varies
during the evolution of the Universe. At the present time the
mass of the neutralino
dark matter particle is given by Eq. (\ref{dmp}) with the scalar field $\phi$
evaluated at the present time $\phi_{0}$. This type of physics
associated with the mass
varying dark matter particles
have been proposed and studied in
the
literature \cite{rosenfeld, hof, mass}, but in these studies the masses of the dark matter
particle are constant in space. In this paper we consider the case that
neutralino masses vary as a function of space, for instance the neutralino
mass in the Sun differs from that evaluated on the cosmological scale.

The effective potential of the dark energy scalar at the inner of the Sun
or the Earth is given by
\begin{eqnarray}
\label{veff}
V_{eff}=\rho_B(\phi) +\rho_\chi(\phi)+V(\phi)\ ,
\end{eqnarray}
where $\rho_B(\phi)$ is the mass density of the baryon matter and
$\rho_\chi(\phi)=n_\chi M_\chi(\phi)$. Here it's straightforward to obtain
$\rho_B(\phi)=\rho_{B_{0}}-\frac{ \lambda_B \rho_{B_{0}} \phi}{m_{\rm p}}$
from Eq. (\ref{phiqq}) in terms of the mass density of the baryon matter
$\rho_{B_{0}}$ in the absence of the interaction (\ref{phiqq}).
We will show later that, by choosing an appropriate parameter $\lambda_B$,
$\rho_B(\phi)\simeq\rho_{B_{0}}$ is a good approximation
in the numerical
discussion.

The value of $\phi$ in the Sun can be obtained by requiring
$\phi$ be at the minimum of the local potential
\begin{equation}
\left. \label{mini} \frac{dV_{eff}}{d\phi}\right|_{\phi=\phi_{min}}=0\ .
\end{equation}
The contributions to the equation above from the dark matter and
baryon are proportional to
$\lambda_B\rho_B$ and $\lambda_\chi\rho_\chi$ respectively.
Since the $\rho_\chi$ is about 15 orders of magnitude smaller than
$\rho_B$ in the core of the Sun \cite{dm}, the influence
of the dark matter on the effective potential of the dark energy scalar
can be safely ignored if
\begin{equation}
\lambda_B \gg \lambda_\chi \frac{\rho_\chi}{\rho_B}
 \sim 10^{-15} \lambda_\chi,
\end{equation}
which we will show can be satisfied generally.
Given Eq. (\ref{mini}) we have for the value of $\phi$
\begin{equation}
\label{phimin}
\phi_{min}=\frac{m_p}{\beta}{\rm ln}\frac{\lambda_B \rho_B}{\beta V_0}\ .
\end{equation}
From Eqs. (\ref{dmp}) and (\ref{phimin}) we obtain
the ratio of the masses of the dark matter particles in the Sun to that
evaluated on the cosmological scale
\begin{equation}
\label{mass}
\frac{M_{\chi}^{\odot}}{M_{\chi}^{cos}}
=\frac{1+\frac{\lambda_\chi}{\beta} \ln
\frac{\lambda_B\rho_B^{\odot} }{\beta V_0}}
{1+\lambda_\chi\frac{\phi_0}{m_p}}\ .
\end{equation}

In Eq. (\ref{mass}) $\phi_0$ is the cosmological value of the scalar field
at the present time. To satisfy the cosmological observations on
the dark energy our numerical results show that $\phi_0 =
-0.51m_p$, $V_0=4.2\times 10^{-47}{\rm GeV}^4$ and
$\beta=1$ and $\lambda_\chi=0.1$ which we have mentioned above.

 The baryon energy density of
the Sun $\rho_B^{\odot}$ is
about $2.5~g/{\rm cm}^3$.
However due to the logarithm in
Eq. (\ref{phimin}), the effect on the dark matter mass is insensitive to
the
baryon mass density $\rho_B$. For example, the baryon density in the Sun
$\rho_B^{\odot}$ is about a quarter of that in the Earth,
which causes that $M_{\chi}^{\odot}$ is about 2\% smaller than
the value at the Earth in our numerical calculation for $\lambda_B =
10^{-9}$.

The parameter $\lambda_B$ is constrained by the tests on the gravitational
inverse square law \cite{inverse}
and the tests on the equivalence principle \cite{equiva} to be
$\lambda_B \lesssim \mathcal{O}(10^{-2})$.
Here we point out that the presence of the
interaction of the dark energy with the baryon in Eq. (\ref{phiqq})
makes the baryon mass density also vary
\begin{equation}
\label{rhob}
\frac{\delta\rho_B}{\rho_B}=\frac{\rho_B(\phi_{min})-\rho_{B_0}}{\rho_B}
=-\frac{\lambda_B}{\beta}{\rm ln} \frac{\lambda_B \rho_{B}}{\beta V_0}.
\end{equation}
If taking $\rho_B(\phi_{min})$ to be the baryon density in the Earth
(which is similar to the baryon density of the Sun),
$\delta\rho_B(\phi_{min})$ indicates the correction of the dark energy
to baryon mass in the Earth.
The proton mass has been measured very precisely on the Earth with an
error
of $10^{-8}$. If we take as an example that
$\delta\rho_B/\rho_B < 10^{-8}$ we obtain an upper
limit on $\lambda_B$, $\lambda_B < 10^{-9}$, which we use in the numerical
calculation.

\begin{figure}
\includegraphics[scale=0.40]{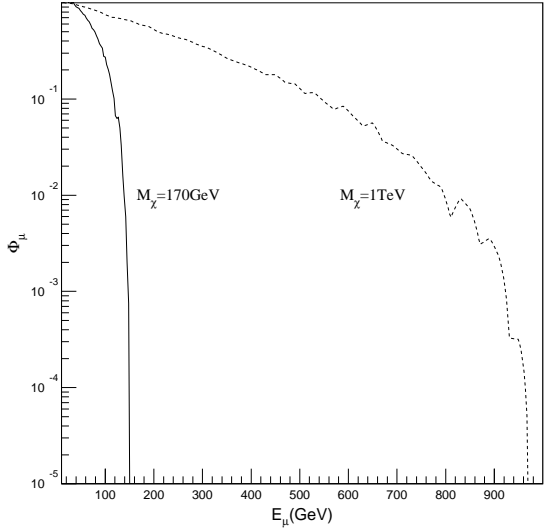}
\caption{\label{fig1}
Spectra of the muons induced by the neutrinos from the annihilation of the neutralino
dark matter particles in the core of the Sun with $m_\chi=1 $ TeV and in the
cosmological scale with $m_\chi= 170 $ GeV.
The spectra are normalized at 10 GeV and calculated by
using the DarkSUSY package \cite{darksusy}.
}
\end{figure}

Now we have $\phi_{min}\simeq 49 m_p$ and consequently
$M_{\chi}^{\odot}/M_{\chi}^{cos}\simeq
6$.
In Fig. \ref{fig1} we plot the muon spectra induced by the neutrino from the dark matter
annihilation in the center of the Sun and in the cosmological scale.
We choose the dark matter mass in the Sun to be 1 TeV, while the dark
matter mass in the cosmological scale is about 1TeV/6=170 GeV. From the figure one can
see clearly the difference in the neutrino spectra.

On the cosmological scale the dark energy scalar is homogeneously distributed,
however in this case it is inhomogeneous, which gives rise to energy
density $\rho_\phi$ in the Sun.
From Eqs. (\ref{poten}) and (\ref{phimin}) we have
\begin{equation}
\frac{\rho_\phi}{\rho_B}\simeq \frac{V(\phi_{min})}{\rho_B}
=\frac{\lambda_B}{\beta}\ .
\end{equation}
For  $\lambda_B < 10^{-9}$ and $\beta=1$
we have $\rho_\phi/\rho_B \sim 10^{-9}$ which shows that the dark energy
density inside the Sun (or the Earth) can be safely ignored.

In summary, we have in this paper studied the possible effects of
interacting dark
energy on the detection of the dark matter particles.
We have taken a specific model of Quintessence with an
exponential potential and interacting with the dark matter mass term,
and then discussed numerically the influence of the dark
energy on the neutrino spectrum from the annihilation of the neutralino
pairs in the Sun.
Our results show the possibility of probing for the dark energy in the
future experiments of searching for the dark matter particles.

\begin{acknowledgments}
We thank Y. Chen, B. Feng and C. Huang for discussions.
This work is supported in part by the NSF of China
under the grant No. 10105004,10120130794,19925523, 90303004
and also by the Ministry of Science and Technology of China under
grant No. NKBRSF G19990754.
\end{acknowledgments}

\end{document}